\documentclass[aps,prl,reprint,floatfix,showpacs]{revtex4-1}

\usepackage{amsmath,amsfonts,amssymb}
\usepackage{graphicx}
\usepackage{color}
\def\beq{\begin{equation}}
\def\eeq{\end{equation}}
\def\beqn{\begin{eqnarray}}
\def\eeqn{\end{eqnarray}}

\begin{document}

\title{Overlap of exact and Gross-Pitaevskii wavefunctions in Bose-Einstein condensates of dilute gases}

\author{Shachar Klaiman}
\affiliation{Theoretische Chemie, Physikalisch--Chemisches Institut, Heidelberg University, 
Im Neuenheimer Feld 229, D-69120 Heidelberg, Germany}
\author{Lorenz S. Cederbaum}
\affiliation{Theoretische Chemie, Physikalisch--Chemisches Institut, Heidelberg University, 
Im Neuenheimer Feld 229, D-69120 Heidelberg, Germany}
\date{\today}

\begin{abstract}
It has been proven theoretically for bosons with two-body repulsive interaction potentials in the dilute limit that the Gross-Pitaevskii equation provides the exact energy and density per particle as does the basic many-particle Schr\"{o}dinger equation [Lieb and Seiringer, Phys. Rev. Lett. 88, 170409 (2002)]. Here, we investigate the overlap of the Gross-Pitaevskii and exact ground state wavefunctions. It is found that this overlap is always smaller than unity and may even vanish in spite of the fact that both wavefunctions provide the same energy and density per particle. Consequences are discussed. 
\end{abstract}


\maketitle

Since the experimental discovery of Bose-Einstein condensates (BECs) consisting of dilute atomic gases two decades ago \cite{BEC1,BEC2,BEC3}, there has been vast interest in their properties \cite{rev1,rev2,rev3}. 
In the respective theoretical studies, the Gross-Pitaevskii equation which is obtained by minimizing the Gross-Pitaevskii energy functional \cite{Yngvason_PRA} has played a particularly leading role. The simplicity of this mean field equation adds much to its popularity as it can be solved rather straightforwardly and exhibits many interesting and appealing properties. Importantly, it has been rigorously proven by Lieb and Seiringer (LS theorem) \cite{Lieb_PRL} that in the dilute limit the Gross-Pitaevskii (GP) equation provides the exact energy and density per particle as does the full many-particle Schr\"{o}dinger equation. One immediate and highly relevant consequence of this proof is that BECs are $100\%$ condensed in the limit of infinite particle number.

In the dilute limit, also called GP limit, the interaction parameter $\Lambda=\lambda_0(N-1)$ appearing in the GP equation, where $\lambda_0$ is the two-particle interaction strength, is kept fixed as $N\rightarrow\infty$. 
 The LS theorem might raise the impression that the GP theory correctly describes BECs with large particle numbers at zero temperature.  Nevertheless, it is well known that corrections beyond the GP theory can be relevant for experiments with typical particle numbers \cite{jackson}. Does GP theory also provide an accurate wavefunction of BECs in the dilute limit? This is a relevant question as, after all, the wavefunction contains all the physical properties of the system.  A first clear indication that boson correlations not included in GP theory can be relevant has been shown very recently by Klaiman and Alon \cite{VAR1,VAR2}. To answer the latter question we have chosen the overlap of the GP and exact ground states as an obvious measure of the quality of the GP wavefunction. The proof by Lieb and Seiringer is restricted to 3 and 2 dimensions and assumes the existence of a finite scattering length, but we would like to go beyond and consider the general case of a many-boson Hamiltonian and its mean field (which we call GP) in the dilute limit. We shall show that the LS theorem applies also for cases not covered by the available proof.

As a first step we introduce a many-body perturbation theory (MBPT) where the unperturbed Hamiltonian is the GP one. The situation is similar to the so called M{\o}ller-Plesset MBPT widely and successfully employed in electronic structure calculations which is based on the Hartree-Fock unperturbed Hamiltonian \cite{MBbook}. The GP Hamiltonian $h_{GP}({\bf r}) = h + v$, where $h$ comprises the kinetic energy of a boson and its trap potential and $v = \Lambda \left|\varphi_{GP}({\bf r})\right|^2$, possesses a complete set of eigenfunctions $\varphi_i({\bf r})$ of which the one with the lowest eigenvalue $\mu_{GP}$ called the chemical potential is $\varphi_{GP}({\bf r})$. The eigenvalue  equation of the GP operator reads:
\beq
\label{eq::1}
\left[h + v\right] \varphi_i({\bf r})=\mu_i\varphi_i({\bf r}).
\eeq
We note that the GP equation can also be derived from c-field methods \cite{blakie}, but being interested here in the wavefunction of the system at zero-temperature we rely here on the quantum mechanical origin of this equation. We may now rewrite the many-body Hamiltonian of N interacting bosons
$H = \sum_{j=1}^Nh\left({\bf r}_j\right)+ \sum_{j>k}\lambda_0 V\left({\bf r}_j-{\bf r}_k\right)$ 
where $V\left({\bf r}_j-{\bf r}_k\right)$ is the boson-boson interaction potential and $\lambda_0$ its strength to give:
\beq
\label{eq::2}
H = H_0 + \lambda_0W.
\eeq
Here, $H_0 = \sum_{j=1}^Nh_{GP}\left({\bf r}_j\right)$   and   $\lambda_0W = \lambda_0V - v$  are now the unperturbed Hamiltonian and the residual interaction suitable for our MBPT. 

The orthonormal eigenstates of $H_0$ can all be cast into the simple form
\beq
\label{eq::3}
\left|q_1,q_2,\ldots,q_m\right\rangle =   \frac{(a_1^\dag)^{q_1}(a_2^\dag)^{q_2}\cdots(a_m^\dag)^{q_m}}{\sqrt{q_1! q_2! \cdots q_m!}}\left|0\right\rangle, 
\eeq
where the $a_i^\dag$ are the usual boson creation operators corresponding to the solutions $\varphi_i\left({\bf r}\right)$ in Eq. \ref{eq::1}, $\left|0\right\rangle$ is the boson vacuum, and the total number of bosons $q_1 + q_2 + \ldots + q_m = N$.  Identifying $a_1^\dag$ with $a_{GP}^\dag$, the N boson  GP ground state is just
$\left|GP\right\rangle = \left|N\right\rangle = [N!]^{-1/2} (a_{GP}^\dag)^N\left|0\right\rangle$.
Note that zero occupations $q_i = 0$  are not indicated in the eigenstates. It is easily seen that
$H_0 \left|q_1,\ldots,q_m\right\rangle = \sum_i\mu_i q_i \left|q_1,\ldots,q_m\right\rangle $  and, in particular, $H_0\left|N\right> =  N\mu_{GP} \left|N\right\rangle$ . 

We are now in the position to write down the relevant MBPT expansion. As can be found in text books \cite{MBbook}, the exact eigenfunction $|\tilde{\Psi} \rangle$ in the intermediate normalization $\langle N | \tilde{\Psi}\rangle = 1$ can be expanded in orders of perturbation 
\begin{align}
\label{eq::4}
|\tilde{\Psi} \rangle &= \sum_{n=0}|\tilde{\Psi}^{(n)} \rangle, \\
\nonumber |\tilde{\Psi}^{(n)} \rangle &= \left\{\frac{\hat{Q}}{N\mu_{GP} - H_0} (\lambda_0W -\Delta E)\right\}^n  |N\rangle.
\end{align}
Here, $|\tilde{\Psi}^{(0)}\rangle = |N\rangle$, $\hat{Q} = 1 - |N\rangle\langle N|$ is a projection operator which removes $|N\rangle$ from the terms $|\tilde{\Psi}^{(n)}\rangle$, $n>0$, and $\Delta E = E_{exact} - N\mu_{GP}$ is the difference between the exact energy and that of the unperturbed Hamiltonian. This increment can also be expanded as
\begin{align*}
\Delta E &= \sum_{n=1}E^{(n)} \mbox{  ,  } E^{(n)} = \langle N| \lambda_0W| \tilde{\Psi}^{(n-1)}\rangle.
\end{align*}
Obviously, $E^{(0)} + E^{(1)}$, where $E^{(0)} = N\mu_{GP}$, is nothing but the total GP energy $E_{GP} = \langle N|H|N\rangle$ of the N boson system. 

The normalized exact many-body ground state is, of course, given by $|\Psi_{exact}\rangle = |\tilde{\Psi} \rangle / \langle \tilde{\Psi}|\tilde{\Psi} \rangle^{1/2}$ , and hence the overlap $S(N)$ between the GP and exact ground states simply takes on the form
\beq
S(N) = \langle GP|\Psi_{exact}\rangle = \langle \tilde{\Psi}|\tilde{\Psi} \rangle^{-1/2} 
\eeq
and because of the projector $\hat{Q}$, we see that
\begin{align}
\label{eq::5}
S(N) &= (1 +  \langle\Delta \Psi |\Delta \Psi\rangle)^{-1/2} \mbox{  ,  } |\Delta \Psi\rangle &=  \sum_{n=1}|\tilde{\Psi}^{(n)}\rangle.
\end{align}
Clearly, this overlap is smaller than $1$. 

Let us now evaluate S(N) in the leading order of perturbation theory which should be valid for small values of the interaction parameter $\Lambda$. We will focus on the dilute limit $N \rightarrow \infty$ and $\Lambda$ kept fixed. To compute any term $|\tilde{\Psi}^{(n)}\rangle$ one inserts in Eq. \ref{eq::4} the unity operator $\hat{1} = \sum|q_1,q_2,\ldots\rangle\langle q_1,q_2,\ldots|$. Being interested in the leading term $|\tilde{\Psi}^{(1)}\rangle$, one immediately sees that only $|N-1,1_i\rangle$, $|N-2,1_i,1_j\rangle$ and $|N-2,2_i\rangle$ unperturbed states with $i,j>1$ contribute. Now, due to the choice $\lambda_0W = \lambda_0V-v$ it can be shown that the matrix elements $\langle N-1,1_i|\lambda_0V|N\rangle$ cancel those of $-v$ in the residual interaction, and we are left only with the latter two kinds of states and their matrix elements of $V$ only. The general rules to evaluate matrix elements of operators in the basis of the Fock states (\ref{eq::3}) can be found in \cite{MCHB}. The final result correct up to second order reads: 
\beq
\label{eq::6}
S(N) = [1 + \Lambda^2\alpha^2]^{-1/2}  \mbox{  for  } N \rightarrow \infty ,
\eeq
where $\alpha^2 = \sum_{i,j}\frac{V_{ij11}^2}{(2\mu_{GP}-\mu_i-\mu_j)^2}(1-\frac{1}{2}\delta_{ij})$ and the matrix element $V_{ij11} = \int\varphi_i({\bf r})\varphi_j({\bf r}')V({\bf r}-{\bf r}')\varphi_{GP}({\bf r})\varphi_{GP}({\bf r}')d{\bf r}d{\bf r}'$. If we evaluate $\alpha^2$ for a one-dimensional case of bosons in a box with a contact interaction $V(x-x') = \delta(x-x')$, one finds that this quantity is essentially the size of the box. Supported by our examples we assume that generally $\alpha^2$ reflects the space available for the bosons, the more space is available the smaller the overlap S is. From Eq. \ref{eq::6} we conclude that independently of how small the interaction parameter $\Lambda$ is, the overlap $S(N)$ is smaller than $1$ also in the dilute limit.

How small can the overlap $S$ become? The general expression (\ref{eq::5}) and the perturbative (\ref{eq::6}) indicate that the overlap will typically drop for increasing interaction parameter $\Lambda$. However, it is difficult to rigorously derive an expression for the overlap at large values of $\Lambda$, and we therefore take recourse to examples. We present two examples, one which is analytically solvable in all dimensions and one which can only be numerically solved. Before presenting our analytical example we stress that for very large boson number $N$ and fixed $\Lambda$, the interaction strength $\lambda_0$ of a pair of bosons is extremely small and proportional to $1/N$. This makes clear that large values of $\Lambda$ can easily be achieved still keeping $\lambda_0$ vanishingly small. In other words, it is absolutely legitimate, theoretically anyway, but also experimentally to consider large values of the interaction parameter $\Lambda$.%

An analytically solvable model of $N$ interacting bosons in a trap exists which is very valuable in discussing the overlap $S(N)$ at large values of $\Lambda$ in the dilute limit. In this model the trap is harmonic and the interaction potential too. We may call it the harmonic interaction model (HIM). This model has been solved explicitly \cite{HIM_Cohen} and investigated in several scenarios \cite{HIM_Yan,HIM_Po1,HIM_Benchmarks,LR_MCTDHB_Bench,Schilling_FER_HIM_WORK,Axel_MCTDHF_HIM}. The many-boson Hamiltonian reads
\begin{equation*}
H=\sum _{i=1}^N (\frac{\hat{{\bf p}}_i^2}{2}+\frac{\omega ^2}{2}{\bf r}_i^2)+\lambda_0\sum _{i<j}^N \left({\bf r}_i-{\bf r}_j\right){}^2,
\end{equation*}
where we use units in which $\hbar=m=1$. The corresponding Schr\"{o}dinger equation is solved by introducing normal coordinates

\begin{equation*}
{\bf Q}_k=\frac{1}{\sqrt{k (k+1)}}\sum _{i=1}^k \left({\bf r}_{k+1}-{\bf r}_i\right) \mbox{  for  } 1<k<N-1
\end{equation*}
and the center of mass coordinate
\begin{equation*}
{\bf Q}_N=\frac{1}{\sqrt{N}}\sum _{i=1}^N {\bf r}_i \mbox{  .}
\end{equation*}
The exact ground state wavefunction of $H$ takes on the appearance 
\begin{align*}
\Psi _{\text{exact}}\left({\bf Q}_1,{\bf Q}_2,\ldots ,{\bf Q}_N\right)&=\left(\frac{\omega }{\pi }\right)^{D/4} \left(\frac{\delta _N}{\pi }\right)^{\frac{D (N-1)}{4}} e^{-\frac{\text{$\omega ${\bf Q}}_N^2}{2}} \\
&\times e^{-\frac{1}{2} \delta _N \sum _{k=1}^{N-1} {\bf Q}_k^2},
\end{align*}
where $D$ is the dimension of the problem (the D=3 result has been reported in \cite{HIM_Cohen}) and the relevant parameter $\delta_N^2 = \omega^2 + 2\lambda_0N$ which for large $N$ becomes $\delta_N^2 = \omega^2 + 2\Lambda$. The GP wavefunction can also be expressed by the above coordinates and takes on the simple appearance \cite{HIM_Cohen}:
\begin{equation*}
\Psi _{\text{GP}}\left({\bf Q}_1,{\bf Q}_2,\ldots ,{\bf Q}_N\right)=\left(\frac{\delta _{N-1}}{\pi }\right)^{\frac{D N}{4}} e^{-\frac{1}{2} \delta _{N-1} \sum _{k=1}^N {\bf Q}_k^2}
\end{equation*}

We have computed explicitly the overlap of these two functions as a function of $N$, $\omega$ and $\lambda_0$. For the sake of brevity we present here the result for large $N$ with $\Lambda$ kept fixed:
\begin{align}
\label{eq::7a}
\tag{7a}
S(N) = 2^{D/2} \frac{\left(  1+ \frac{2 \Lambda}{\omega ^2}\right)^{D/8}}{\left(1+\sqrt{1+\frac{2 \Lambda}{\omega ^2}}\right)^{D/2}}
\end{align}
For small $\Lambda$ one readily obtains 
\begin{align}
\label{eq::7b}
\tag{7b}
S(N)= (1+\Lambda^2\alpha^2)^{-1/2} + \mathcal{O}((\Lambda\alpha)^3) \mbox{\quad;  }  \alpha^2=\frac{D}{8\omega^2}
\end{align}
which demonstrates how the overlap decreases as the dimension of the trap increases and also when the ``size'' of the trap ($\sim 1/\omega$) increases.  More importantly, we are now in the position to see what happens for large $\Lambda$, where perturbation theory, of course, does not apply. From (\ref{eq::7a}) one immediately gets:
\begin{align}
\label{eq::7c}
\tag{7c}
S(N)=2^{3D/8}\left(\frac{\Lambda}{\omega^2}\right)^{-D/8}
\end{align}

Obviously, the overlap between the GP and the exact ground state wavefunctions approaches zero as $\Lambda$ becomes large, and, interestingly, the faster the larger is the dimension of the problem. We stress that in the HIM model the energy and density (also density matrix) per particle in the dilute limit are exactly reproduced by the GP theory \cite{HIM_Cohen}.

\begin{figure}[!]
\includegraphics[width=1\columnwidth,angle=0]{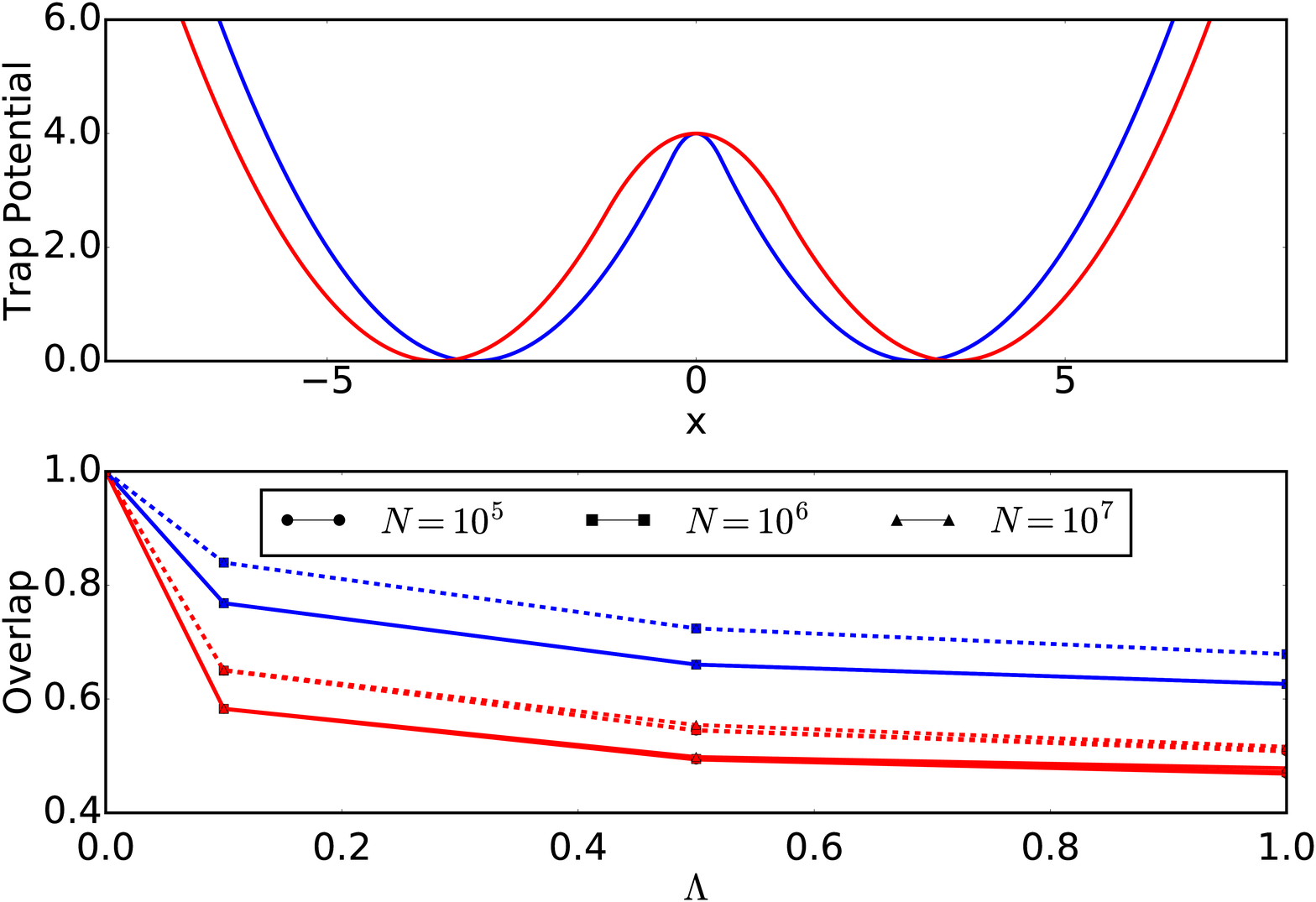}
\caption{(Color online) The overlap of the GP and the numerically computed many-body wavefunctions in the case of 1D and 2D double-well trap potentials. Upper panel: The two trap potentials used in the calculations (in the 2D example $\frac{1}{2}y^2$ has to be added). Lower panel: The overlap as a function of the interaction parameter  for three different boson numbers N. Note that for each value of $\Lambda$  the points computed for different values of N fall on top of each other. The 1D and 2D results are connected by solid and dashed lines, respectively.}
\label{fig1}
\end{figure}

We would like to also study examples with short range interactions. In the absence of exactly solvable models we have to resort to a numerical solution of the full Schr\"{o}dinger equation which is not an easy task for large boson numbers. We investigate a one-dimensional (1D) double-well trap potential and contact interaction $V(x-x') = \delta(x-x')$, a problem widely studied in the literature \cite{Milburn_1997,Smerzi_1997,Smerzi_1999,
Kasevich_2001,Vardi_2001,Markus_2005,Joerg_2005,Jeff_2007,Posa_2009,BJJ,Markus_2010,Universality,Shmuel_2015}. Since the current proof of the LS theorem does not cover 1D, we also extend this example to 2D by choosing V to be a normalized Gaussian, see \cite{Doganov,Fischer}. Two trap potentials are studied, see upper panel of Fig. \ref{fig1}. The trap potential is constructed by connecting two harmonic potentials $V_\pm(x) = \frac{1}{2}(x \pm x_0)^2$ with a cubic spline such that the resulting barrier is of a given height $V_0$. For the 2D example we add the harmonic trap $\frac{1}{2}y^2$. The mass of the particles is chosen to be $1$ as in the case of the HIM investigated above. We variationally solve the Schr\"{o}dinger equation by using the multi-configurational time-dependent Hartree for bosons (MCTDHB) method \cite{MCTDHB1,MCTDHB2} finding the ground state by imaginary time propagation. The MCTDHB is a well established method in the literature \cite{book_MCTDH}. In principle, it is a numerically exact method \cite{HIM_Benchmarks}, but for large boson numbers it can only be solved approximately as the number of boson Fock states fiercely explodes. If MCTDHB is used with a single variational single-particle function, say $\varphi_g$, the working equations boil down to give exactly the GP equation, i.e., $\varphi_g = \varphi_{GP}$. If, on the other hand, we use two variational single-particle functions (called orbitals), say $\varphi_g$ and $\varphi_u$, then the many-body state obtained becomes a superposition of $N+1$ Fock states and reads
\beq
\label{eq::8}
\tag{8}
|\Psi\rangle=\sum_{n=0}^NC_n|N-n,n\rangle,
\eeq
where the first entry $N-n$ refers to the number of bosons residing in the gerade orbital $\varphi_g$ and $n$ residing in the second, ungerade, orbital $\varphi_u$. In MCTDHB, the orbitals and the coefficients $C_n$ are determined from the time-dependent variational principle \cite{MCTDHB1,MCTDHB2}. Numerically, we find that as the boson number $N$ grows, the gerade orbital $\varphi_g$ smoothly approaches the GP one $\varphi_{GP}$. This finding is very useful, as the overlap $S(N)$ can be simply computed from the first coefficient in (\ref{eq::8}):  $S(N) = C_0$. The ungerade orbital is, however, found to be different from the ungerade solution of the GP equation (\ref{eq::1}). 

We could solve the MCTDHB with two orbitals for up to $N = 10^7$ bosons. The results for the overlap are shown in the lower panel of Fig. \ref{fig1} for three particle numbers and are similar for 1D and 2D. It is clearly seen that the overlap drops as the interaction parameter grows from $\Lambda=0$ to $\Lambda=1$ (in all calculations $\omega=1$). Although $\Lambda$ is rather moderate, the overlap can fall below $0.5$. We would like to stress that the results shown seem to saturate as N is increased: the curves for $N = 10^5$, $10^6$ and $10^7$ essentially fall on top of each other. In other words, the dilute limit is essentially achieved in this example. For the evolution of the overlap from few particles to 10 million particles for one value of $\Lambda$, see supplemental material. One also sees that changes in the trap potential are reflected in the value of the overlap. The wider trap leads to smaller overlaps. 

\begin{figure}[!]
\includegraphics[width=1.0\columnwidth,angle=0]{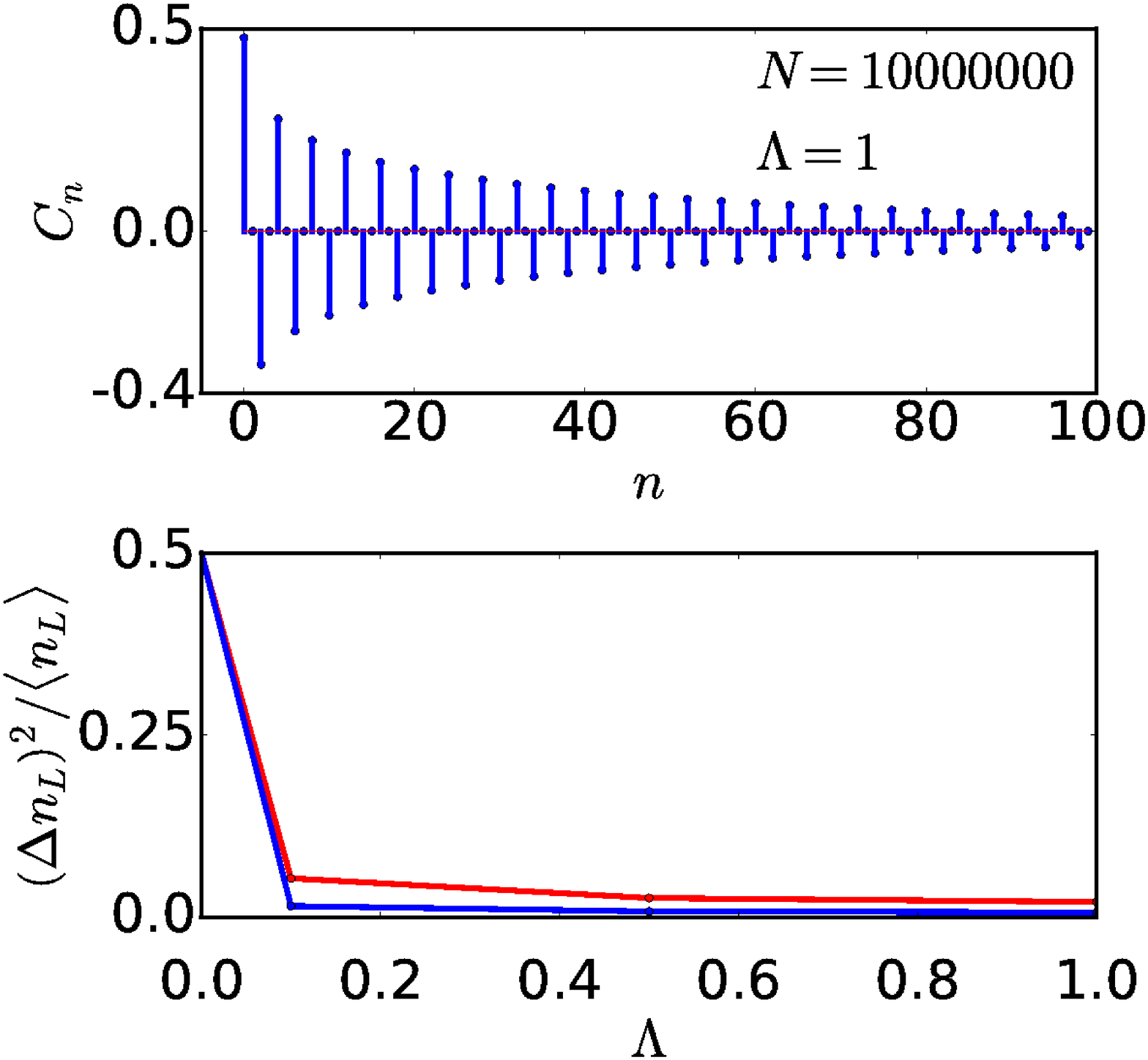}
\caption{(Color online) The impact of boson correlation on the boson number fluctuation in one of the wells of the double-well one-dimensional trap potentials shown in the upper panel of Fig. \ref{fig1} .Upper panel: The expansion coefficients $C_n$ of the many-body wavefunction in \eqref{eq::8}. Note that the overlap $S(N)$ depicted in Fig. \ref{fig1} is provided by the first coefficient $C_0$. Lower panel: The boson number fluctuations relative to the number $N/2$ of bosons in the left well as a function of $\Lambda$ for the two double-well potentials. Note that in GP theory this relative boson number fluctuation is always 0.5. In all calculations $N = 10^7$. }
\label{fig2}
\end{figure}

Having the rather involved and highly correlated wavefunction (\ref{eq::8}) at our disposal, we can compute more involved quantities which reflect the boson correlations. The coefficients $C_n$ are shown for one calculation in 1D in the upper panel of Fig. \ref{fig2}. Although, their distribution is qualitatively extremely different from those dictated by GP ($C_0 =1$ , $C_n = 0$ for $n>0$) the energy and density (also density matrix) per particle in our example coincide numerically very well with the respective GP results (see also the supplemental material for 1D and 2D). This is {\it a posteriori} an interesting finding: a highly complex wavefunction and a one-term wavefunction give the same results.  

Having a double-well trap, we calculate also the particle number fluctuation in one well, say the left well L. This can be done by introducing the creation operators $a_g^\dag = (a_L^\dag + a_R^\dag)/\sqrt{2}$ and $a_u^\dag = (a_L^\dag - a_R^\dag)/\sqrt{2}$ corresponding to the orbitals $\varphi_g$ and $\varphi_u$ which define the left and right orbitals localized in the respective wells (see, e.g., \cite{Sakmann_fluc}). The boson number fluctuation is as usual described by 
\beq
\label{eq::9}
\tag{9}
(\Delta n_L)^2 = \langle \Psi|(a_L^\dag a_L)^2|\Psi\rangle - (\langle \Psi|a_L^\dag a_L|\Psi\rangle)^2 . 
\eeq
In GP theory ($\Psi=\Psi_{GP}$) the resulting number fluctuation is given by: $(\Delta n_L)^2 = N/4$.  Obviously, the number of bosons in one well is just $N/2$. 

Using the correlated wavefunction \eqref{eq::8}, the formal result reads
\begin{align}
\label{eq::10}
\tag{10}
(\Delta &n_L)^2 = \frac{N}{4}+\frac{1}{2}\sum_{n=0}^{N} C_n^2(N-n)n \\
\nonumber &+\frac{1}{2}\sum_{n=2}^{N}C_nC_{n-2}\sqrt{(N-n+1)(N-n+2)n(n-1)}.
\end{align}
Our numerical results for 1D are depicted in the lower panel of Fig. \ref{fig2}. Surprisingly, the boson number fluctuations decrease dramatically with increasing interaction parameter $\Lambda$. The finding rather reminds of a Mott insulator than of a superfluid \cite{rev1,rev2,rev3,Sakmann_fluc}, although the system is essentially condensed. The results are the more surprising if one notices that adding even a single boson outside of the $N-1$ GP bosons enhances the number fluctuation by a factor of $3$. Generally, $(\Delta n_L)^2 = (2n+1) N/4$ if computed with $|\Psi\rangle = |N-n,n\rangle$, for large $N$. Obviously, the cross terms in (10) are those which are responsible for the substantial suppression of the boson number fluctuations.

In the GP limit in which $N \rightarrow\infty$ and the interaction parameter $\Lambda$ is kept fixed, the total energy as well as the density per boson are exactly reproduced by the GP theory. Nevertheless, we find that the overlap of the GP and exact many-body wavefunctions is always smaller than $1$, and depending on the trap and $\Lambda$, can be rather small, even vanishingly small. This in turn implies that the exact wavefunction describes substantial boson correlations, by definition not present in GP theory. Obviously, the energy and density per boson are mean field quantities in the GP limit. The situation is very different from that in fermion systems, e.g., in electronic systems like atoms and molecules, where the respective mean field theory is Hartree-Fock. Since two fermions cannot occupy the same one-particle state (orbital), they build up a shell structure and their total energy does not depend only on one interaction parameter and fermion correlations are reflected in the total energy. 

Although we find it very interesting that in the dilute limit GP provides the energy and density per boson correctly even if the overlap of the GP wavefunction with the exact one can be essentially zero, one clearly does not catch the rich many-particle physics present in condensed boson systems by studying GP theory or by measuring energy and density. Indeed, this overlap behavior tells us that the underlying many-particle physics is rich. Other, boson-correlation susceptible quantities should be computed and measured. One example is the boson number fluctuation discussed here, but there are many other. We refer, for instance, to the recently proposed single-shot measurements which contain much information on the system beyond mean field \cite{KasparNat} and predictions of the effect of correlations in the GP limit on many-body variances \cite{VAR1,VAR2}. 

Finally, we would like to briefly remark on excited states and dynamics. In the ground state GP theory provides the lowest energy per particle in the dilute limit. In excited states other mean field (called best mean field \cite{BMF}) functionals can provide lower energy than GP even in the dilute limit. An example can be found in \cite{BMF1} where the system is not condensed but exhibits a macroscopic fragmentation. Time-dependent GP theory is also often employed to compute the dynamics of a system. Even if one starts the process with a condensed state, it is clear that excited many-particle states will mix in as time proceeds and as boson correlation is expected to be more present in excited states, deviations from GP are expected to grow in time \cite{BJJ,Universality}. Thus, investigating fragmentation and boson-correlation susceptible quantities in dynamical processes may show the deviations from mean field theory more clearly for large boson numbers. 
%



\end{document}